\documentclass[11pt,a4paper]{article}
\usepackage[hyperref]{acl2021}
\usepackage{times}
\usepackage{latexsym}

\usepackage{verbatim}
\usepackage{graphicx}
\usepackage{xcolor}
\usepackage{caption}
\usepackage{subcaption}
\usepackage{amsmath}
\usepackage{amssymb}
\usepackage{enumitem}
\usepackage{multirow}
\usepackage{longtable}

\usepackage{color,soul}
\usepackage{microtype}

\aclfinalcopy 

\title{Understanding and Countering Stereotypes: A Computational Approach to the Stereotype Content Model}

 \author{Kathleen C. Fraser, Isar Nejadgholi, and Svetlana Kiritchenko \\
  National Research Council Canada \\
  Ottawa, Canada \\
 \footnotesize \texttt{\{Kathleen.Fraser,Isar.Nejadgholi,Svetlana.Kiritchenko\}@nrc-cnrc.gc.ca}\\
 }

\date{}

\begin{document}
\maketitle
\begin{abstract}

Stereotypical language expresses widely-held beliefs about different social categories.  Many stereotypes are overtly negative, while others may appear positive on the surface, but still lead to negative consequences.  
In this work, we present a computational approach to interpreting stereotypes in text through the Stereotype Content Model (SCM), a comprehensive causal theory from social psychology. The SCM proposes that stereotypes can be understood along two primary dimensions: warmth and competence. We present a method for defining warmth and competence axes in semantic embedding space, and show that the four quadrants defined by this subspace accurately represent the warmth and competence concepts, according to annotated lexicons. We then apply our computational SCM model to textual stereotype data and show that it compares favourably with survey-based studies in the psychological literature. Furthermore, we explore various strategies to counter stereotypical beliefs with anti-stereotypes. It is known that countering stereotypes with anti-stereotypical examples is one of the most effective ways to reduce biased thinking, yet the problem of generating anti-stereotypes has not been previously studied. 
Thus, a better understanding of how to generate realistic and effective anti-stereotypes can contribute to addressing pressing societal concerns of stereotyping, prejudice, and discrimination.

\end{abstract}

\section{Introduction}

Stereotypes are widely-held beliefs about traits or characteristics of groups of people. While we tend to think of stereotypes as expressing negative views of groups, 
some stereotypes actually express positive views (e.g. \textit{all women are nurturing}). However, even so-called `positive’ stereotypes can be harmful, as they dictate particular roles that individuals are expected to fulfill, regardless of whether they have the ability or desire to do so \cite{Kay2013}.
 
The existence of stereotypes in our society -- including in entertainment, the workplace, public discourse, and even legal policy -- can lead to a number of harms. \citet{Timmer2011} organizes these harms into three main categories: 
(1)~Misrecognition effects: harms caused by denying members of particular groups an equal place in society, diminishing their human dignity, or other forms of marginalization. (2)~Distribution effects: harms resulting from unfair allocation of resources, either by increasing the burden placed on a group, or decreasing a group’s access to a benefit. (3)~Psychological effects: the distress and unhappiness caused by an awareness and internalization of the stereotyped biases against one’s identity group. 
Additionally, the internalization of these negative stereotypes can lead to anxiety and underachievement. 
To reduce these harms and promote a more egalitarian society, we must identify and counter stereotypical language when it occurs.

Evidence from the psychological literature suggests that one of the most effective methods for reducing stereotypical thinking is through exposure to counter-stereotypes, or anti-stereotypes. \citet{Finnegan2015} showed participants stereotypical and anti-stereotypical images of highly socially-gendered professions (e.g., a surgeon is stereotypically male, and a nurse is stereotypically female; the genders were reversed in the anti-stereotypical images), 
and then measured their gender bias in a judgement task. Exposure to anti-stereotypical images significantly reduced gender bias on the task.  \citet{Blair2001} used a mental imagery task 
and reported that participants in the anti-stereotypical condition subsequently showed significantly weaker effects on the Implicit Association Test (IAT). \citet{Dasgupta2001} showed 
a similar effect by exposing participants to anti-stereotypical exemplars (e.g. admired Black celebrities, and disliked white individuals).
When \citet{Lai2014} compared 17 interventions aimed at reducing stereotypical thinking, methods involving anti-stereotypes were  most successful overall.
 
Thus, creating technology that enables users to identify 
stereotypical language when it occurs, and then counter it with anti-stereotypes, could help to reduce biased thinking. However, the idea of what constitutes an anti-stereotype remains ill-defined. Is an anti-stereotype simply the semantic opposite of a stereotype? Or can anything that is not a stereotype serve as an anti-stereotype? If two groups are stereotyped similarly, do they have an identical anti-stereotype? Can an anti-stereotype actually reflect an equally harmful view of a target group (e.g. \textit{the cold-hearted career woman} as an anti-stereotype to \textit{the nurturing housewife)}?
 
Here, we begin to untangle some of these questions using
the StereoSet dataset \cite{Nadeem2020}. We begin by analyzing the stereotypes expressed in this dataset. One widely-accepted model of 
stereotypes, prejudice, and inter-group relationships from social psychology is the ``Stereotype Content Model'' or SCM \citep{Fiske2002}. The SCM proposes two fundamental and universal 
dimensions of social stereotypes: \textit{warmth} and \textit{competence}.  By defining the warm--cold, competent--incompetent axes in the semantic embedding space, we are able to cluster and interpret stereotypes with respect to those axes. We can then examine the associated anti-stereotypes and their relation to both the stereotyped description and the target group. Thus, our contributions are as follows:

\begin{itemize}[topsep=2pt,leftmargin=*]
\itemsep 0em 
\item To develop a computational method for automatically mapping textual information to the warmth--competence plane as proposed in the Stereotype Content Model.
\item To validate the computational method and optimize the choice of word embedding model using a lexicon of words known to be associated with positive and negative warmth and competence.
\item To compare the stereotypes in StereoSet with those reported in the survey-based social psychology literature.
\item To analyze human-generated anti-stereotypes as a first step towards automatically generating anti-stereotypes, as a method of countering stereotypes in text with constructive, alternative perspectives. 
\end{itemize}

\section{Related Work}

We provide more details on the Stereotype Content Model and its practical implications, and then briefly review the NLP research on computational analysis of stereotypical and abusive content.

\noindent\textbf{Stereotype Content Model}:  
Stereotypes, and the related concepts of prejudice and discrimination, have been extensively studied by psychologists for over a century \cite{dovidio2010sage}. 
Conceptual frameworks have emerged which emphasize two principle dimensions of social cognition. 
The Stereotype Content Model (SCM) refers to these two dimensions as \textit{warmth} 
(encompassing sociability and morality) and \textit{competence} 
(encompassing ability and agency) \cite{Fiske2002}. 
When forming a cognitive representation of a social group to anticipate probable behaviors and traits, people are predominantly concerned with the others' intent---are they friends or foes? 
This intent is captured in the  primary dimension of warmth. 
The competence dimension determines if the others are capable to enact that intent. 
A key finding of the SCM has been that, in contrast to previous views of prejudice as a uniformly negative attitude towards a group, many stereotypes are actually \textit{ambivalent}; that is, they are high on one dimension and low on the other. 

Further, the SCM proposes a comprehensive causal theory, linking stereotypes with  social structure, emotions, and discrimination \cite{fiske2015intergroup}. 
According to this theory, stereotypes are affected by a perceived social structure of \textit{interdependence} (cooperation versus competition), corresponding to the warmth dimension, and \textit{status} (prestige and power), determining competence. 
Stereotypes then predict emotional response or prejudices. 
For example, groups perceived as unfriendly and incompetent (e.g., homeless people, drug addicts) evoke disgust and contempt, groups allegedly high in warmth but low in competence (e.g., older people, people with disabilities) evoke pity, and groups perceived as cold and capable (e.g., rich people, businesspeople) elicit envy. 

Finally, the emotions regulate the actions (active or passive help or harm). 
Thus, low warmth--low competence groups 
often elicit active harm and passive neglect, whereas low warmth--high competence groups 
may include envied out-groups who are subjects of passive help in peace times but can become targets of attack during social unrest \cite{cuddy2007bias}. 

The SCM has been supported by extensive quantitative and qualitative analyses across cultures and time \cite{fiske2015intergroup,Fiske2016cultures}. To our knowledge, the current work presents the first computational model of the SCM. 

\noindent\textbf{Stereotypes in Language Models}: An active line of NLP research is dedicated to quantifying and mitigating stereotypical biases in language models. Early works focused on gender and racial bias and revealed stereotypical associations and common prejudices present in word embeddings through association tests \cite{bolukbasi2016man,caliskan2017semantics,manzini-etal-2019-black}. To discover stereotypical associations in contextualized word embeddings, \citet{may-etal-2019-measuring} and \citet{kurita-etal-2019-measuring} used pre-defined sentence templates. 
Similarly, \citet{bartl-etal-2020-unmasking} built a template-based corpus to quantify bias in neural language models, whereas \citet{Nadeem2020} and \citet{nangia2020crows} used crowd-sourced stereotypical and anti-stereotypical sentences for the same purpose. 
In contrast to these studies, while we do use word embeddings to represent our data, we aim to identify and categorize stereotypical views expressed in text, not in word embeddings or language models.

\noindent\textbf{Abusive Content Detection:} 
Stereotyping, explicitly or implicitly expressed in communication, can have a detrimental effect on its target, and can be considered a form of abusive behavior.
Online abuse, including hate speech, cyber-bullying, online harassment, and other types of offensive and toxic behaviors,  
has been a focus of substantial research effort in the NLP community in the past decade
(e.g.\@ see surveys by \citet{schmidt2017survey,fortuna2018survey, vidgen-etal-2019-challenges}).
Most of the successes in identifying abusive content have been reported on text containing explicitly obscene expressions;
only recently has work started on identifying more subtly expressed abuse, such as stereotyping and micro-aggressions \cite{breitfeller-etal-2019-finding}. 
For example, \citet{Fersini2018} and \citet{Chiril2020} examined gender-related stereotypes as a sub-category of sexist language, and \citet{price-etal-2020-six} annotated `unfair generalizations' as one attribute of unhealthy online conversation. 
\citet{sap-etal-2020-social} employed large-scale language models in an attempt to automatically reconstruct stereotypes implicitly expressed in abusive social media posts. 
Their work showed that while the current models can accurately predict whether the online post is offensive or not, they struggle to effectively reproduce human-written statements for implied meaning.

\vspace{2mm}
\noindent\textbf{Counter-narrative:} Counter-narrative (or counterspeech) has been shown to be effective in confronting online abuse \cite{benesch2016counterspeech}. 
Counter-narrative is a non-aggressive response to abusive content that aims to deconstruct and delegitimize the harmful beliefs and misinformation with thoughtful reasoning and fact-bound arguments. 
Several datasets of counter narratives, spontaneously written by regular users or carefully crafted by experts, have been collected and analyzed to discover common intervention strategies \cite{mathew2018analyzing,chung-etal-2019-conan}. 
Preliminary experiments in automatic generation of counter-narrative demonstrated the inadequacy of current large-scale language models for generating effective responses and the need for a human-in-the-loop approach \cite{qian-etal-2019-benchmark,tekiroglu-etal-2020-generating}. 
Countering stereotypes through exposure to anti-stereotypical exemplars is based on a similar idea of deconstructing harmful beliefs
with counter-facts.

\section{Data and Methods}

We develop our computational SCM using labelled data from \citet{Nicolas2020} and the POLAR framework for interpretable word embeddings \cite{mathew2020polar}, and then apply it to stereotype and anti-stereotype 
data from StereoSet \cite{Nadeem2020}. Details are provided in the following sections.

\subsection{Warmth-Competence Lexicons}

To construct and validate our model, we make use of the supplementary data from \citet{Nicolas2020} (\url{https://osf.io/yx45f/}). They provide a list of English seed words, captured from the psychological literature, associated with the warmth and competence dimensions; specifically, associated with sociability and morality (warmth), and ability and agency (competence).
They then use WordNet to generate an extended lexicon of English words either positively or negatively associated with aspects of warmth and competence.
Some examples from the seed data and extended lexicon are given in Table~\ref{tab:seed_words}.

\begin{table*}[]
    \centering
    \small
    \begin{tabular}{l l l l r l  r }
    \hline 
    Dimension & Component & Sign & Seed word examples & $n_{\mathrm{seed}}$ & Extended lexicon examples & $n_{\mathrm{extended}}$ \\
    \hline
     Warmth & Sociability & pos & friendly, warm, pleasant & 34 & amusing, brother, fun & 482 \\
            &     & neg & cold, repellent, disliked & 32 & detached, grim, surly & 423\\
     & Morality    & pos & trustworthy, sincere, honest & 40 & donor, justice, modest &460 \\
            &      & neg & dishonest, selfish, unfair & 49 & cheat, dreadful, henchman & 1750\\
    \hline
     Competence & Agency      & pos & confident, assertive, secure & 35 & bravery, decisive, stubborn & 444\\
     &             & neg & fearful, lazy, inactive & 31 & follow, minion, quitter & 265\\
     & Ability     & pos & smart, intelligent, able & 33 & analytic, fluency, thorough & 579\\
     &             & neg & stupid, ignorant, incapable & 29 & forgetful, silly, unfit & 301\\
    \hline
    \end{tabular}
    \caption{Examples of words from the training data (seed words) and validation data (extended lexicon), for each of the components comprising the warmth and competence dimensions.}
    \label{tab:seed_words}
\end{table*}

\subsection{StereoSet}

For human-generated stereotype and anti-stereotype data, we use the publicly-available portion of the StereoSet dataset \cite{Nadeem2020}. This English-language dataset was constructed to test language model bias, and part of the data is kept hidden as the test set for a leaderboard on language model fairness (\url{https://stereoset.mit.edu/}). 
Instead, we use the development set, which contains stereotype data for 79 target groups across four broad demographic domains: gender, race or nationality, profession, and religion. 

In StereoSet, there are two experimental conditions: intra-sentence and inter-sentence. Here, we focus on the intra-sentence data only. The data was collected from crowd-workers as follows (see \citet{Nadeem2020} for more detail): Given a target group label, the annotator is asked to generate a stereotypical word associated with that group, as well as an anti-stereotypical word and an unrelated word. They then construct a context sentence containing the target group label, and a blank which can be filled with the stereotypical or anti-stereotypical word. For example, if the target group was women, the annotator might come up with \textit{emotional} and \textit{rational} as the stereotype and anti-stereotype words respectively, and then construct a sentence like \textit{Women are known for being overly $\langle$BLANK$\rangle$}. 
For our current analysis, we consider only the stereotype and anti-stereotype words, and discard the context sentence. We also exclude any targets that do not directly refer to groups of people (e.g., we discard \textit{Norway} but keep \textit{Norwegian}).  This results in 58 target groups with an average of 25 stereotype and anti-stereotype word pairs each.

\subsection{Constructing Warmth and Competence Dimensions}

We consider several possible representations for the words in our dataset, including GloVe \cite{Pennington2014glove}, word2vec \cite{Mikolov2013efficient}, and FastText \cite{mikolov2018advances}.\footnote{We consider here only noncontextual word embeddings, in line with  \citet{mathew2020polar}. Because the POLAR framework is based on linear algebraic computations, it is not immediately obvious whether it will extend directly to contextualized embeddings, which are notably anisotropic \cite{Ethayarajh2019}.} In all cases, the key question is how to project the higher-dimensional word embedding onto the warmth--competence plane. 

Rather than using an unsupervised approach such as PCA, we choose the POLAR framework introduced by \citet{mathew2020polar}. This framework seeks to improve the interpretability of word embeddings by leveraging the concept of `semantic differentials,' a psychological rating scale which contrasts bipolar adjectives, e.g. \textit{hot--cold}, or \textit{good--bad}. Given word embeddings that define these polar opposites for a set of concepts, all other word embeddings in the space are projected onto the `polar embedding space,' where each dimension is clearly associated with a concept.

For our purposes, the polar opposites are warmth--coldness and competence--incompetence, as defined by the sets of seed words from \citet{Nicolas2020}. To reduce the dimensionality of the space to 2D, we average the word vectors for all seed words associated with each dimension and polarity. That is, 
to define the warmth direction, we take the mean of all words in the seed dictionary which are positively associated with warmth. Given vector definitions for warmth, coldness, competence, and incompetence, we can then use a simple matrix transformation to project any word embedding to the 2D subspace defined by these basis vectors (mathematical details are given in Appendix A).

\section{Model Validation}

We first evaluate the model's ability to accurately place individual words from the lexicons along the warmth and competence dimensions. We then explore whether we can reproduce findings describing where certain target groups are typically located in the warmth--competence plane, based on the previous survey-based social psychology literature. 

\subsection{Comparison with Existing Lexicons}

As described above, we use the extended lexicon from \citet{Nicolas2020} to validate our model. We remove any words in the lexicon which appear in the seed dictionary and any words which do not have representations in all the pretrained embedding models, leaving a total of  
3,159 words for validation.

In the extended lexicon, the words are annotated with either +1 or -1 to indicate a positive or negative association with the given dimension. We pass the same words through our system, and observe whether the model labels the word as being positively or negatively associated with the relevant dimension. Our evaluation metric is accuracy; i.e.\@ the proportion of times our system agrees with the lexicon.
Note that all words are associated with \textit{either} warmth or competence, and therefore we can only evaluate one dimension at a time. 

We evaluate a number of pre-trained word embeddings in the gensim library \cite{rehurek2010}, with the results given in Table~\ref{tab:validation}. The FastText embeddings generally outperform the other embeddings on this task, with the 2M word model trained on 600B tokens in the Common Crawl leading to the highest accuracy. Therefore, we use this embedding model in the analysis that follows.

\begin{table}[t!]
    \centering
    \small
    \begin{tabular}{l c c}
    \hline
    Embedding model & Warmth & Comp.\\
    \hline 
    FastText-crawl-subword-300 & \textbf{85.0} & \textbf{85.8} \\
    FastText-wiki-news-subword-300 & 84.9 & 84.8 \\
    Word2vec-GoogleNews-300 & 80.2 & 72.6\\
    GloVe-twitter-200 & 72.8 & 74.2\\
    GloVe-wiki-gigaword-300 & 78.7 & 77.9 \\

\hline
    \end{tabular}
    \caption{Accuracy of the word embedding models on predicting the correct labels for the extended lexicon.}
    \label{tab:validation}
\end{table}

\subsection{Comparison with Psychological Surveys}

We now address the question of whether our model, in conjunction with the StereoSet data, is able to reproduce findings from psychological surveys. 
We project stereotypes from the StereoSet data onto the warmth--competence space for the 24 target groups that meet both of the following criteria: (1) they are included in the publicly available portion of the StereoSet data, and (2) they have been previously studied for stereotyping in the psychological literature.
Based on the findings from psychological surveys, we expect these target groups will be mapped to the following quadrants:\footnote{Note that these research findings simply report stereotypical beliefs which are prevalent in North American society; we in no way aim to perpetuate, confirm, or promote these views.}

\begin{itemize}[topsep=10pt,leftmargin=*]
\itemsep 0em 
\item \textbf{Warm-Competent:} nurse, psychologist (`healthcare professions') \citep{Brambilla2010}, researcher ('professor') \cite{Eckes2002}.

\item \textbf{Warm-Incompetent:} grandfather (`elderly'), mommy, mother (`traditional women') \citep{Cuddy2008}, schoolboy, schoolgirl (`children') \citep{Fiske2018}.

\item \textbf{Cold-Competent:} male, gentleman (`man') \citep{Glick2004}, Japanese \citep{Lee2006}, commander \citep{Cuddy2011}, manager, entrepreneur \citep{Fiske2010}, mathematician, physicist, chemist, engineer (`scientist') \cite{Losh2008}, software developer (`technical expert') \cite{Fiske2018}.

\item \textbf{Cold-Incompetent:} African, Ethiopian, Ghanian, Eritrean, Hispanic \citep{Lee2006}, Arab \citep{Fiske2006}.
\end{itemize}

\noindent To locate each target group on the plane, we  generate word embeddings for each of the stereotype words associated with the target group, find the mean, and project the mean to the polar embedding space. As we aim to identify commonly-held stereotypes, we use a simple cosine distance filter to remove outliers, heuristically defined here as any words which are greater than a distance of 0.6 from the mean of the set of words.
We also remove words which directly reference a demographic group (e.g., black, white) as these words are vulnerable to racial bias in the embedding model and complicate the interpretation. 
A complete list of the words in each stereotype cluster can be found in the Appendix B.

\begin{figure}[t!]
         \centering
         \includegraphics[width=0.5\textwidth]{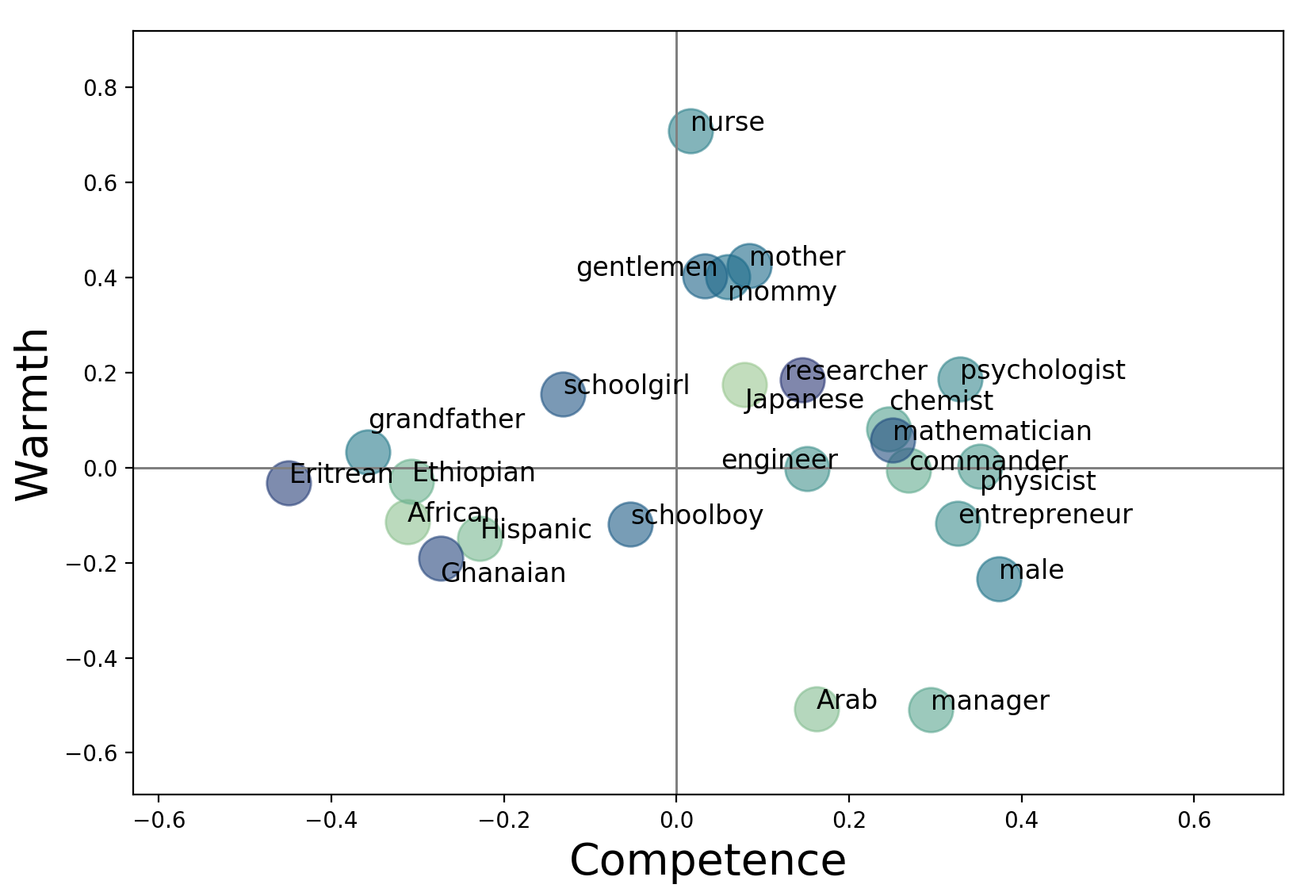}
         \caption{Validating known stereotypes.  }
         \label{fig:stereotypes_validation}
\end{figure}

Figure~\ref{fig:stereotypes_validation} confirms many of the findings predicted by the literature. Most (67\%) of the stereotypes lie in the predicted quadrant, including \textit{grandfather} and \textit{schoolgirl} in the paternalistic warm--incompetent quadrant; \textit{nurse} and \textit{psychologist} in the admired warm--competent quadrant, \textit{manager} and \textit{male} in the envied cold--competent quadrant, and \textit{African} and \textit{Hispanic} in the cold--cold quadrant. 

Other stereotypes lie in locations which seem reasonable on examination of the underlying data. For example, while men are typically stereotyped as being competent yet cold in the psychological literature, the specific keyword \textit{gentlemen} evokes a certain subset of men (described with words such as \textit{polite}, \textit{respectful}, and \textit{considerate}), which ranks higher on the warmth dimension than the target word \textit{male} (example words: \textit{dominant}, \textit{aggressive}).

We also observe that while children have generally been labelled as warm--incompetent in previous work \citep{Fiske2018}, this dataset distinguishes between male and female schoolchildren, and, as expected based on studies of gender, schoolboys are ranked as lower warmth  than schoolgirls. The words used to describe schoolboys include references to the `naughty' schoolboy stereotype, while the words describing schoolgirls focus on their innocence and naivety. 

It is also notable that \textit{Arab}, predicted to lie in the cold--incompetent quadrant, is here mapped to the cold--competent quadrant instead. We hypothesize that this is due to the use of stereotype words like \textit{dangerous} and \textit{violent}, which suggest a certain degree of agency and the ability to carry out goals. In contrast, the target group \textit{African} as well as those associated with African countries are stereotyped as \textit{poor} and \textit{uneducated}, and thus low on the competence dimension.

In general, we conclude that in most cases the computational approach is successful in mapping stereotyped groups onto the predicted areas of the warmth--competence plane, and that the cases which diverge from findings in the previous literature do appear to be reasonable, based on an examination of the text data. Having validated the model, we can now apply it to the rest of the stereotype data in StereoSet, as well as the anti-stereotypes. 

\begin{table*}[tbh]
    \centering
    \small 
    \begin{tabular}{l | r | r r r r | r r }
    \hline
    Strategy          & Overall        & HC-HW              & LC-HW              & LC-LW             & HC-LW             & $|$C$| > |$W$|$  & $|$W$| > |$C$|$\\
                      & $n = 895$       & $n = 192$         & $n = 183$     & $n = 176$     & $n = 344$ & $n = 428$ & $n = 467$ \\

    \hline
    
    Direct antonym    & 23.4           & 26.0            & \textbf{32.6 }  & 27.8            & 15.0           & 27.2          & 19.2\\
    Opposite quadrant & \textbf{29.6}  & \textbf{30.2}   & 15.5            & 26.1            & \textbf{38.3}  &\textbf{28.1} &\textbf{31.2}\\
    Flip warmth       & 20.6           & 14.6            & 26.5            & \textbf{29.5 }  & 16.4           & 12.3          & 29.8\\
    Flip competence   & 16.7           & 24.0            & 12.7            & 13.1            & 16.7           & 22.8          & 10.1\\
    Same quadrant     & 9.6            & 5.2              & 12.7           & 3.4             &   13.5         & 9.6           & 9.6\\
    \hline
    \end{tabular}
    \caption{The percentage of times each of the hypothesized strategies of anti-stereotype generation is used for stereotypes, overall and in each quadrant. Quadrants are labelled as HC-HW, LC-HW, LC-LW, and HC-LW, where HC/LC denotes high/low competence, and HW/LW denotes high/low warmth. 
    We also consider separately those stereotypes which have competence as the most salient dimension ($|$C$| > |$W$|$), and those which have warmth as the most salient dimension ($|$W$| > |$C$|$). }
    \label{tab:antistereotypes}
\end{table*}

\section{Stereotypes and Anti-Stereotypes}
\label{sec:antistereotypes}

The SCM presents a concise theory to explain stereotypes and resulting prejudiced behaviour; however, it does not generate any predictions about anti-stereotypes. Here, we explore the anti-stereotypes in StereoSet within the context of the SCM, first at the level of individual annotators, and then at the level of target groups (combining words from multiple annotators).
We then discuss how we might use information about warmth and competence to generate anti-stereotypes with the specific goal of reducing biased thinking.

\subsection{Anti-Stereotypes in StereoSet}

In this section, we investigate the question: What do human annotators come up with when asked to produce an anti-stereotype? One possibility is that they simply produce the antonym of their stereotype word. To test this hypothesis, for all 58 groups and each pair of stereotype and anti-stereotype words, we obtain a list of antonyms for the stereotype word using the Python library PyDictionary. 
We additionally search all the synonyms for the stereotype word, and add all of their antonyms to the list of antonyms as well. Then, if the lemma of the anti-stereotype matches the lemma of any of the retrieved antonyms, we consider it a match. 

However, as seen in Table~\ref{tab:antistereotypes}, the strategy of simply producing a direct antonym is only used 23\% of the time. We consider four other broad possibilities: (1) that the annotator generates an anti-stereotype word that lies in the opposite quadrant from the stereotype word, e.g., if the stereotype word is low-competence, low-warmth (LC-LW), then the anti-stereotype word should be high-competence, high-warmth (HC-HW); (2) that the annotator chooses a word with the opposite warmth polarity (i.e. flips warmth), while keeping the competence polarity the same; (3) that the annotator chooses a word with the opposite competence polarity (i.e. flips competence), while keeping the warmth polarity the same; (4) that the annotator chooses a word that lies in the same quadrant as the stereotype word. We report the proportion of times that each strategy is observed; first overall, then for each quadrant individually. The choice of whether to modify warmth or competence might also depend on which of those dimensions is most salient for a given word, and so we consider separately words for which the absolute value of competence is greater than the absolute value of warmth, and vice versa. The results are given in Table~\ref{tab:antistereotypes}.

While no single strategy dominates, we can make a few observations. In general, it is more likely that people select an anti-stereotype which is not a direct antonym, but which lies in the opposite quadrant in the warmth-competence plane. 
Flipping only one axis is less frequent, although we see in the last two columns that it is more likely that the competence will be flipped when competence is the salient dimension for a word, 
and similarly for warmth.
Finally, choosing another word in the same quadrant is rare, but more common in the ambivalent quadrants.

While it is not possible to know what thought process the annotators followed to produce anti-stereotypes, we consider the following possible explanation. Just as we have here conceptualized a stereotype as being  defined not by a single word, but by a set of words, perhaps each annotator also mentally represents each stereotype as a set of words or ideas. Then, the anti-stereotype word they produce sometimes reflects a different component of their mental image than the initial stereotype word. To give a concrete example from the data, one annotator stereotypes Hispanic people as \textit{aggressive}, but then comes up with \textit{hygienic} as an anti-stereotype, suggesting that \textit{unhygienic} is also part of their multi-dimensional stereotype concept. The choice of whether to select a direct antonym, or whether to negate some other component of the stereotype, may depend on the availability of a familiar lexical antonym, the context sentence, or any number of other factors. In short, it appears that the process by which human annotators generate pairs of stereotype and anti-stereotype words is complex and not easily predicted by the SCM.

We then examine how these pairs of stereotype and anti-stereotype words combine to produce an overall anti-stereotype for the target group in question. Taking the same approach as in the previous section, we average the anti-stereotype word vectors to determine the location of the anti-stereotype in the warmth--competence plane. 
For each target group, we then select the word closest to the mean for both the stereotype and anti-stereotype clusters. Similarly to when we look at individual word pairs, in 22\% of cases, the mean of the anti-stereotype is the direct antonym of the stereotype mean. In the other cases, 45\% of the anti-stereotype means lie in the opposite quadrant to the stereotypes, in 16\% of cases the warmth polarity is flipped, in 10\% of cases the competence polarity is flipped, and in only 7\% cases (4 target groups), the anti-stereotype lies in the same quadrant as the stereotype.

\begin{table}[]
    \centering
    \small 
    \begin{tabular}{p{1.8cm} p{1.2cm} p{1.1cm} p{2.0cm}}
    \hline
   Target& Stereotype & Antonym & Anti-stereotype \\ 
   \hline
    African & poor & rich & rich \\
    Hispanic & poor & rich & hardworking \\
    mother & caring & uncaring & hateful \\
    nurse & caring & uncaring & rude \\
    commander & strong & weak & stupid \\
    mover & strong & weak & weak \\
    football player & dumb & smart & weak \\
    \hline 
    \end{tabular}
    \caption{Examples comparing stereotypes with their direct antonym and the anti-stereotype from StereoSet. 
    }
    \label{tab:antonyms}
\end{table}

In Table~\ref{tab:antonyms}, we offer a few examples of cases where the anti-stereotype means agree and disagree with the direct antonyms of the stereotypes. As in the pairwise analysis, in many cases the anti-stereotypes appear to be emphasizing a supposed characteristic of the target group which is not captured by the stereotype mean; for example, the anti-stereotype for `dumb football player' is not \textit{smart}, but \textit{weak} -- demonstrating that \textit{strength} is also part of the football player stereotype. This is also seen clearly in the fact that two target groups with the same stereotype mean are not always assigned the same anti-stereotype: for example, both Africans and Hispanics are stereotyped as \textit{poor}, but Africans are assigned the straightforward anti-stereotype \textit{rich}, while Hispanics are assigned \textit{hard-working} (perhaps implying that their poverty is due to laziness rather than circumstance). 

The general conclusion from these experiments is that stereotypes are indeed multi-dimensional, and the anti-stereotypes must be, also. Hence it is not enough to generate an anti-stereotype simply by taking the antonym of the most representative word, nor is it sufficient to identify the most salient dimension of the stereotype and only adjust that. When generating anti-stereotypes, annotators (individually, in the pairwise comparison, and on average) tend to invert both the warmth and competence dimensions, taking into account multiple stereotypical characteristics of the target group.

\subsection{Anti-Stereotypes for Social Good}

The anti-stereotypes in StereoSet were generated with the goal of evaluating language model bias. Ultimately, our goal is quite different: to reduce biased thinking in humans. 
In particular, we want to generate anti-stereotypes that emphasize the positive aspects of the target groups. 

\begin{table*}[h]
    \centering
    \small
    \begin{tabular}{l p{1.5cm} l l l}
    \hline
    Target & Stereotype  & Anti-stereotype  & \textit{X but Y }construction & \textit{X and $\neg$Y} anti-stereotype\\
    \hline 
      Grandfather  & old & young & kind but feeble & kind and strong \\
      Entrepreneur & savvy & lazy & inventive but ruthless & inventive and compassionate \\
      Engineer     & smart & dumb & intelligent but egotistical & intelligent and altruistic \\
      Mommy        & loving & uncaring & caring but childish & caring and mature \\
      Software developer & nerdy & dumb & intelligent but unhealthy & intelligent and healthy \\
      \hline
    \end{tabular}
    \caption{Examples of positive anti-stereotypes created by identifying positive and negative words along each of the dimensions, and taking the antonym only of the negative words. }
    \label{tab:social_good}
\end{table*}

As underscored by \citet{Cuddy2008}, many stereotypes are ambivalent: they take the form 'X but Y'. Women are \textit{nurturing but weak}, scientists are \textit{intelligent but anti-social}. When we simply take the antonym of the mean, we focus on the single most-representative word; i.e., the X. However, in the examples we can observe that it's actually what comes after the ``but~...'' that is the problem. Therefore, in generating anti-stereotypes for these ambivalent stereotypes, we hypothesize that a better approach is not to take the antonym of the primary stereotype (i.e., women are \textit{uncaring}, scientists are \textit{stupid}), but rather to challenge the secondary stereotype (women can be nurturing \textit{and strong}, scientists can be intelligent \textit{and social}). 

As a first step towards generating anti-stereotypes for such ambivalent stereotypes, we propose the following approach: first identify the most positive aspect of the stereotype (e.g., if the stereotype mean lies in the incompetent--warm quadrant, the word expressing the highest warmth), then identify the most negative aspect of the stereotype in the other dimension (in this example, the word expressing the lowest competence). Then the stereotype can be phrased in the X but Y construction, where X is the positive aspect and Y is the negative aspect.\footnote{A similar method can be used for warm--competent and cold--incompetent stereotypes, although if all words are positive, an anti-stereotype may not be needed, and if all words are negative, then an antonym may be more appropriate.} 
To generate a positive anti-stereotype which challenges stereotypical thinking while not promoting a negative view of the target group, take the antonym only of the negative aspect. Some examples are given in Table~\ref{tab:social_good}. A formal evaluation of these anti-stereotypes would involve carrying out a controlled psychological study in which the anti-stereotypes were embedded in an implicit bias task to see which formulations are most effective at reducing bias; for now, we simply present them as a possible way forward.

As shown in the table, taking into account the ambivalent aspects of stereotypes can result in more realistic anti-stereotypes than either taking the mean of the crowd-sourced anti-stereotypes, or simply generating the semantic opposite of the stereotype. For example, the group \textit{grandfather}  is mostly stereotyped as \textit{old}, and then counter-intuitively anti-stereotyped as \textit{young}. 
It is more useful in terms of countering ageism to combat the underlying stereotype that grandfathers are \textit{feeble} rather than denying that they are often old. Similarly, it does not seem helpful to oppose biased thinking by insisting that entrepreneurs can be \textit{lazy}, engineers and developers can be \textit{dumb}, and mothers can be \textit{uncaring}. Rather, by countering only the negative dimension of ambivalent stereotypes, we can create realistic and positive anti-stereotypes.

\section{Discussion and Future Work}
\label{sec:discussion}

Despite their prevalence, stereotypes can be hard to recognize and understand. We tend to think about other people on a group level rather than on an individual level because
social categorization, although harmful, simplifies the world for us and leads to cognitive ease. 
However, psychologists have shown that we can overcome such ways of thinking with exposure to information that contradicts those biases.
In this exploratory study, we present a computational implementation of the Stereotype Content Model 
 to better understand and counter stereotypes in text.

A computational SCM-based framework can be a promising tool for large-scale analysis of stereotypes, by mapping a disparate set of stereotypes to the 2D semantic space of warmth and competence. 
We described here our first steps towards developing and validating this framework, on a highly constrained dataset: in StereoSet, the annotators were explicitly instructed to produce stereotypical ideas, the target groups and stereotypical words are clearly specified, and every stereotype has an associated anti-stereotype generated by the same annotator. 
In future work, this method should be further assessed by using different datasets and scenarios. For example, it may be possible to collect stereotypical descriptions of target groups `in the wild' by searching large corpora 
from social media or other sources. We plan to extend this framework to analyze stereotypes on the sentence-level and consider the larger context of the conversations. 
Working with real social media texts will introduce a number of challenges, but will offer the possibility of exploring a wider range of marginalized groups and cultural viewpoints. 

Related to this, we reiterate that only a portion of the StereoSet dataset is publicly available. Therefore, the data does not include the full set of common stereotypical beliefs for social groups frequently targeted by stereotyping. In fact, some of the most affected communities (e.g., North American Indigenous people, LGBTQ+ community, people with disabilities, etc.) are completely missing from the dataset. 
In this work, we use this dataset only for illustration purposes and preliminary evaluation of the proposed methodology. Future work should examine data from a wide variety of subpopulations differing in language, ethnicity, cultural background, geographical location, and other characteristics.

From a technical perspective, with larger datasets it will be possible to implement a cluster analysis \textit{within} each target group to reveal the different ways in which a given group can be stereotyped. A classification model may additionally improve the accuracy of the warmth--competence categorization, although we have chosen the POLAR framework here for its interpretability and ease of visualization.

We also examined how we might leverage the developed computational model to challenge stereotypical thinking. 
Our analysis did not reveal a simple, intuitive explanation for the anti-stereotypes produced by the annotators,
suggesting they exploited additional information beyond what was stated in the stereotype word.
This extra information may not be captured in a single pair of stereotype--anti-stereotype words, but by considering \textit{sets} of words, we can better characterize stereotypes as multi-dimensional and often ambivalent concepts, consistent with the established view in psychology. This also allows us to suggest anti-stereotypes which maintain positive beliefs about a group, while challenging negative beliefs.

We propose that this methodology may potentially contribute to technology that assists human professionals, such as psychologists, educators, human rights activists, etc., in identifying, tracking, analyzing, and countering stereotypes at large scale in various communication channels.  
There are a number of ways in which counter-stereotypes can be introduced to users (e.g., through mentions of counter-stereotypical members of the group or facts countering the common beliefs) with the goal of priming users to look at others as individuals and not as stereotypical group representatives. 
An SCM-based approach can provide the psychological basis and the interpretation of automatic suggestions to users.

Since our methodology is intended to be part of a technology-in-the-loop approach, where the final decision on which anti-stereotypes to use and in what way will be made by human professionals, we anticipate few instances where incorrect (i.e., not related, unrealistic, or ineffective) automatically generated anti-stereotypes would be disseminated. In most such cases, since anti-stereotypes are designed to be positive, no harm is expected to be incurred on the affected group. However, it is possible that a positive, seemingly harmless anti-stereotypical description can have a detrimental effect on the target group, or possibly even introduce previously absent biases into the discourse. Further work should investigate the efficiency and potential harms of such approaches in real-life social settings.

\section*{Ethical Considerations}

\noindent\textbf{Data:} We present a method for mapping a set of words that represent a stereotypical view of a social category held by a given subpopulation onto the two-dimensional space of warmth and competence. The Stereotype Content Model, on which the methodology is based, has been shown to be applicable across cultures, sub-populations, and time \cite{fiske2015intergroup,Fiske2016cultures}. Therefore, the methodology is not specific to any subpopulation or any target social group.

In the current work, we employ the publicly available portion of the StereoSet dataset \cite{Nadeem2020}. This English-only dataset has been created through crowd-sourcing US workers on Amazon Mechanical Turk. Since Mechanical Turk US workers tend to be younger and have on average lower household income than the general US population \cite{difallah2018demographics}, the collected data may not represent the stereotypical views of the wider population. Populations from other parts of the world, and even sub-populations in the US, may have different stereotypical views of the same social groups. 
Furthermore, as discussed in Section~\ref{sec:discussion}, the StereoSet dataset does not include stereotype data for a large number of historically marginalized groups. Future work should examine data both referring to, and produced by, a wider range of social and cultural groups.

\noindent\textbf{Potential Applications:} 
As discussed previously, the automatically proposed anti-stereotypes can be utilized by human professionals in a variety of ways, e.g., searching for or creating anti-stereotypical images, writing counter-narratives, creating educational resources, etc. One potential concern which has not received attention in the related literature is the possibility that the process of generating counter-stereotypes may itself introduce new biases into the discourse, particularly if these counter-stereotypes are generated automatically, perhaps even in response to adversarial data. We emphasize the importance of using counter-stereotypes \textit{not} to define new, prescriptive boxes into which groups of people must fit (e.g., from Table~\ref{tab:antistereotypes}, that all software developers should be intelligent and healthy, or that all entrepreneurs must be inventive and compassionate). Rather, counter-stereotypes should weaken common stereotypical associations by emphasizing that any social group is not actually homogenous, but a group of individuals with distinct traits and characteristics.  In most cases, the algorithm-in-the-loop approach (with automatic suggestions assisting human users) should be adopted to reduce the risk of algorithmic biases being introduced into the public discourse.

Often, harmful stereotyping is applied to minority groups.
Work on identifying and analyzing stereotypes might propagate the harmful beliefs further, and it is possible that collections of stereotypical descriptions  could be misused as information sources for targeted campaigns against vulnerable populations. However, this same information is needed to understand and counter stereotypical views of  society. We also note that although we take advantage of word embedding models in our approach, we do not use the representations of target group names. Previous work has shown that biased thinking is encoded in these models, and using them to represent groups can be harmful to specific demographics.

\vspace{2mm}
\noindent\textbf{Identifying Demographic Characteristics:} The proposed methodology deals with societal-level stereotypical and anti-stereotypical representations of groups of people and does not attempt to identify individual user/writer demographic characteristics. 
However, work on stereotyping and anti-stereotyping entails, by definition, naming and defining social categories of people. 
Labeling groups not only defines the category boundaries, but also positions them in a hierarchical social-category taxonomy \cite{beukeboom2019stereotypes}. 
We emphasize that our goal is not to maintain and reproduce existing social hierarchies, as cautioned by \citet{blodgett-etal-2020-language}, but rather to help dismantle this kind of categorical thinking through the use of anti-stereotypes.

\vspace{2mm}
\noindent\textbf{Energy Resources:} The proposed SCM-based method 
is computationally low-cost, and all experiments were performed on a single CPU. Once the pretrained vectors are loaded, the projection and analysis is completed in less than a minute.

\bibliographystyle{acl_natbib}
\bibliography{anthology,acl2021}

\clearpage 
\appendix

\section{Constructing POLAR dimensions} 

In contrast to the standard POLAR framework introduced by \citet{mathew2020polar}, we do not have a set of polar opposite word pairs, each representing a different interpretable dimension,  but rather a set of words for each of the concepts warmth, coldness, competence, and incompetence from \citet{Nicolas2020}. 
Therefore, we use a slightly different formulation to obtain the polar directions associated with warmth and competence.\footnote{We use the same notation as \citet{mathew2020polar} to explain our method.}

Let $\mathbb{D} = [\overrightarrow{\mathbb{W}_1^a}, \overrightarrow{\mathbb{W}_2^a}, \overrightarrow{\mathbb{W}_3^a},...,\overrightarrow{\mathbb{W}_V^a}] \in \mathbb{R}^{V \times d}$ denote the set of pretrained $d$-dimensional embedding vectors, trained with algorithm $a$, where $V$ is the size of the vocabulary and $\overrightarrow{\mathbb{W}_i^a}$ is a unit vector representing the $i_{th}$ word in the vocabulary. 

In this work, we use four sets of seed words; a set of $N_1$ words associated with positive warmth $\mathbb{P}_{w+}=\{p_{w+}^1,p_{w+}^2,...,p_{w+}^{N_1}\} $, a set of $N_2$ words associated with negative warmth, $\mathbb{P}_{w-}=\{p_{w-}^1,p_{w-}^2,...,p_{w-}^{N_2}\} $, a set of $N_3$ words associated with positive competence, $\mathbb{P}_{c+}=\{p_{c+}^1,p_{c+}^2,...,p_{c+}^{N_3}\} $, and a set of $N_4$ words associated with negative competence, $\mathbb{P}_{c-}=\{p_{c-}^1,p_{c-}^2,...,p_{c-}^{N_4}\}$. In order to find the two polar opposites, we obtain the following directions:

\renewcommand{\theequation}{A.\arabic{equation}}

\begin{equation}
\begin{split}
    \overrightarrow{dir_1} = \frac{1}{N_1} \sum_{i=1}^{N_1}{\mathbb{W}_{p_{w+}^i}^a} - \frac{1}{N_2} \sum_{i=1}^{N_2}{\mathbb{W}_{p_{w-}^i}^a}\\
    \overrightarrow{dir_2} = \frac{1}{N_3} \sum_{i=1}^{N_3}{\mathbb{W}_{p_{c+}^i}^a} - \frac{1}{N_4} \sum_{i=1}^{N_4}{\mathbb{W}_{p_{c-}^i}^a}
\end{split}
\end{equation}

\noindent where $\mathbb{W}_{\upsilon}^a$  represents the vector of the word $\upsilon$. The two direction vectors are stacked to form $ dir \in \mathbb{R}^{2 \times d}$, which represents the change of basis matrix for the new two-dimensional embedding subspace $\mathbb{E}$. In the new subspace, a word $\upsilon$ is represented by $\overrightarrow{\mathbb{E}}_\upsilon$, which is calculated using the following linear transformation: 

\begin{equation}
\begin{split}
    \overrightarrow{\mathbb{E}}_\upsilon = (dir^T)^{-1}\mathbb{W}_\upsilon^a
\end{split}
\end{equation}
In our experiments, we showed that, as expected, the two dimensions of $\mathbb{E}$ are associated with warmth and competence. 

\section{Stereotype Data}

Here we present all of the words contributing to each stereotype for each target group. In addition to 37 tokens which did not have vector representations in the pre-trained embeddings, there were two reasons why words were discarded from the analysis, as described in the paper. First, if a word directly referenced a demographic category, it was discarded. This was to avoid, as much as possible, including effects of language model bias in our model. For example, a number of annotators used the word \textit{black} to describe prisoners; however, if the language model has some racial bias involving the word \textit{black}, then it would affect the placement of the word \textit{prisoner} on the warmth-competence plane. While acknowledging that stereotypical associations between groups are problematic and worth of study in their own right (including this disturbing example involving race and incarceration), it is beyond the scope of the current analysis. 

Additional words were discarded in a filtering step, where words greater than a cosine distance of 0.6 from the mean or centroid of the group were discarded. As people's views towards different groups naturally vary, this was done to prevent outlier words from impacting the analysis, which is focused here on the most widespread or prevalent stereotypes of a given group. While heuristically chosen, the threshold value appears to be acceptable in many cases (see Table~\ref{tab:stereoset-data}). However, other times a large number of words are discarded, which appear in some cases to represent a second, coherent cluster of stereotype words for a given group (see, for example, the words for \textit{policeman}, which split into two clusters alternately characterizing the group as corrupt and racist, or strong and heroic). As mentioned in the discussion, future work will examine how we can identify clusters of stereotypes in larger datasets. 

In the following table, for each target group we present three lists of words:

\begin{itemize}[noitemsep,topsep=0pt,leftmargin=*]
    \item \textbf{Included:} the words included in the analysis. This list is ranked according to distance to the mean, and thus most `representative' words occur first. Words which occur more than once have the frequency given in parentheses.
    \item \textbf{Discarded:} the words discarded as outliers by the thresholding step.
    \item \textbf{Demographic:} the words discarded as referring primarily to a demographic characteristic. 
\end{itemize}

\clearpage
\onecolumn

\setcounter{table}{0}
\renewcommand{\thetable}{B.\arabic{table}}

\begin{longtable}{ p{1.5cm}| p{13.5cm}}

   \centering
     \textbf{Target group} & \textbf{Words} \\
    \hline
\hline
\hline
    \multicolumn{2}{c}{\textbf{\small{Target groups from the `gender' category}}}\\
\hline
\hline
      \small{mother}  &  \small{\textbf{Included:} \textit{caring(6), loving(4), nurturing(5), comforting, protective, kind(2), love, strong, feminine, overbearing(2), children, housekeeper} \textbf{Discarded :}  \textit{old } \textbf{Demographic:} []}\\
     \hline
      
      \small{mommy} & \small{\textbf{Included:} \textit{loving(4), caring(4), nurturing(5), doting, sweet, protective, understanding, child, sexy, overbearing, mama, childish, busy }\textbf{Discarded:} \textit{kindly, nagging } \textbf{Demographic:} \textit{female }} \\
       \hline
       
       \small{schoolgirl} & \small{\textbf{Included:} \textit{smart(2), naive, immature(2), young, innocent(2), sexy, studious, artsy, girly, small, hopeful, thin, hardworker }      \textbf{Discarded:} \textit{innnocent, short, cellphone, does} \textbf{Demographic:} []} \\
       \hline 
       
       \small{schoolboy} & \small{\textbf{Included:} \textit{naughty(2), mischevious(3), immature(2), innocent, cute, energetic, horny, young, troublesome, brainy, troublemaker, pranks}  \textbf{Discarded:} \textit{nerd, weak, athletic, pretentious, strong, student, touched, mean, football, scholar, tag, uniform} \textbf{Demographic:} []} \\
       \hline 
       
       \small{male} & \small{\textbf{Included:} \textit{strong(8), dominant(3), aggressive, muscular, masculine(2), violent, strength, brute, abusive, stupid, dumb}  \textbf{Discarded:} \textit{blue, messy, ignorant, demanding, alpha, manager, sports, mansplaining } \textbf{Demographic:} []}\\
       \hline

       \small{gentlemen} & \small{\textbf{Included:} \textit{polite(3), respectful, classy(2), considerate, gentle, kind, cordial, honorable, mannerable, handsome, proper(2), haughty, nicest, respect, bossy} \textbf{Discarded:}  \textit{nerdy, open, wealthy, wealthy, understanding, were, flask} \textbf{Demographic:} []} \\
       \hline
       
        \small{herself} & \small{\textbf{Included:} \textit{emotional(2), kind, dumb, negative, upbeat, beautiful, unsure, incapable, mean} \textit{girl} \textbf{Discarded:}  \textit{lonely, uptight, created, dress, pink, slap, cake, beating, shopping, alone, needlepoint, walked} \textbf{Demographic:} []} \\
       \hline
       
       \small{himself} & \small{\textbf{Included:} \textit{athletic, progressive, brawny, strong, runner, workout, egotistical, gaming} \textit{male, boy } \textbf{Discarded:}  \textit{funny, quiet, doctors, safe, job, aggression, admire, yardwork, enjoy, push, opinions, alone} \textbf{Demographic:} []  }\\
       \hline
       
       \small{grandfather} & \small{\textbf{Included:} \textit{old(8), antiquated, feeble, boring, stubborn, kind} \textbf{Discarded:} \textit{wise(4), smart, sexist, slow(2), hero, sleeping, protected, watch, spoil, fight} \textbf{Demographic:} []}\\
       \hline
       
       \small{sister} & \small{\textbf{Included:} \textit{caring(2), loving, nurturing, talkative, sassy, bossy, empathy, fragile, friend} \textbf{Discarded:}  \textit{annoying(2), close, small, bonded, protect, argumentative, naggy, noisy, mean(2), college, catty, rival } \textbf{Demographic:} \textit{girl}}\\
       \hline

\hline
\hline
\multicolumn{2}{c}{\small{\textbf{Target groups from the `religion' category}}}\\
\hline
\hline
      \small{Muslim} & \small{\textbf{Included:} \textit{violent(2), terrorist(6), dangerous(2), evil(2), radical(2), hateful, religious, aggressive, crazy, misogynist, intollerant, oppression, legalistic, prejudice, unprecedented} \textbf{Discarded:}  \textit{religion, quaran, hijab} \textbf{Demographic:} \textit{men}}\\
       \hline
       
        \small{Brahmin} & \small{\textbf{Included:} \textit{religious(3), sacred, ancient, foreign, temples, nobleman, loving } \textbf{Discarded:}  \textit{prideful, snobby, arrogant, haughty, wise, unknown, despised, dirty(2), prestigeous, creator(2)} \textbf{Demographic:}  \textit{Muslim, Hindu} }\\
       \hline
       
\hline
\hline
\multicolumn{2}{c}{\small{\textbf{Target groups from the `race or nationality' category}}}\\
\hline
\hline

       \small{Eritrean} & \small{\textbf{Included:} \textit{poor(6), impoverished, lazy(2), needy, dirty(2), rich, starving, weird, aggressive, unsustainable } \textbf{Discarded:} \textit{proud, gang, crime, submissive, stole, huts, artwork, rug} \textbf{Demographic:} \textit{black(2)}} \\
       \hline
       
       \small{Norweigan} & \small{\textbf{Included:}\textit{blond(7), pale, tall, attractive} \textbf{Discarded:}  \textit{beautiful, wealthy(2), cold(2), strong, serious, nice, sincere, quiet} \textbf{Demographic:} \textit{white(4)}} \\
       \hline
       
        \small{Saudi Arabian} & \small{\textbf{Included:} \textit{rich(4), greedy(2), poor, spoiled, oppressed, conservative } \textbf{Discarded:} \textit{cruel, oil, docile, pretentious, sandy, racist, hateful, prince, mecca, camel, extremist, turbans, terrorist(2), isis } \textbf{Demographic:} \textit{Muslim(4)}} \\
       \hline
       
       \small{Japanese} & \small{\textbf{Included:} \textit{friendly, smart, quiet, polite, playful, peaceful, wise, clean(2), sneaky, loyal, passive, educated, shy, exotic, proud, xenophobic} \textbf{Discarded:} \textit{sexist, short, academic, work, kamikaze, sushi, seafood, workaholics}  \textbf{Demographic:} \textit{black }} \\
       \hline
       
        \small{Russian} & \small{\textbf{Included:} \textit{drunk(4), alcoholic, drinking, angry, stupid, vodka, cold, homophobic} \textbf{Discarded:}  \textit{corrupt, suspicious, scary, outspoken, communist, brutal, spy, deceitful, spies, slavic, shady, strong, tough, stern, mean(2), scammer, control, communism }  \textbf{Demographic:} []}\\
       \hline
       
       \small{Ethiopian}  & \small{\textbf{Included:} \textit{skinny(7), starving(5), hungry(2), emaciated, thin } \textit{Christian}  \textbf{Discarded:} \textit{empty, unfortunate, corrupt, somber, stews, save}  \textbf{Demographic:} []} \\
       \hline
       
       \small{Columbian} & \small{\textbf{Included:} \textit{drugs(8), cocaine(2), druggie, coffee(2), criminal, druglord } \textbf{Discarded:} \textit{football, hot, sexy(2), local, educated, poor, voluptuous, undeveloped, fiery} \textbf{Demographic:} []} \\
       \hline
       
        \small{Crimean} & \small{\textbf{Included:} \textit{poor, impoverished, weak, nonexistent, corrupt, harsh, indigenous, boring, wartorn, native, abundance, drunk} \textbf{Discarded:} radical, communist, exotic, independent, controversial, stolen, combative, crime, war, lawlessness, dumber, Ukraine \textbf{Demographic:} Islam, Russian, black, white} \\
       \hline
       
        \small{Iranian}  & \small{\textbf{Included:} \textit{dangerous(4), violent, scary, hateful, bloodthirsty, evil, foolish, dumb, terrorist(2), religious, pretentious, conservative, rogue, greasy} \textbf{Discarded:}  \textit{terrorists(3), panicked, war, fragrant, traditional, hookah } \textbf{Demographic:} []}\\
       \hline
       
       \small{Persian people}  & \small{\textbf{Included:} \textit{greedy(2), rich(2), uneducated(2), poor, extravagant, disgusting, intelligent, mysterious, flashy} \textbf{Discarded:}  \textit{hairy, smelly, violent, religious, smoke, strict, sand, naan  } \textbf{Demographic:} \textit{black, Muslim}} \\
       \hline
       
       \small{Arab}       & \small{\textbf{Included:} \textit{violent, dangerous(2), evil, terrorist(2), radical, crazy, religious(2), extremism, extreme, bad, cruel, terrorists(2), dirty(2), conservative, misogynist, war, diverse} \textbf{Discarded:}  \textit{rich(2), oil, turban}  \textbf{Demographic:} \textit{Muslim }}\\
       \hline
  
         \small{African} & \small{\textbf{Included:} \textit{poor(3), impoverished, starving(2), dirty, stupid, harsh, dark, fast(2), violent} \textbf{Discarded:}  \textit{savages, criminal, crackhead, stealing, run, atheltic } \textbf{Demographic:} \textit{black(6)}} \\
       \hline

      \small{Ghanaian} & \small{\textbf{Included:} \textit{poor(4), dangerous(2), dark(2), lazy, strange, reckless, brutish, fun, starving  } \textbf{Discarded:}\textit{ friendly,  athletic, criminals, tribal, less, gold, rapists, tall, dated  } \textbf{Demographic:} \textit{black(2), Indian }} \\
      \hline
       
      \small{Ukrainian} & \small{\textbf{Included:} \textit{stupid, dumb, poor(3), dirty, hateful, cynical, evil, unintelligent, pretty, cheap, powerless, unclean, exotic, criminal, interesting } \textbf{Discarded:} \textit{ religious, isolationist, quiet, freezing, gangster, discernible, vodka(3), allies, borscht }\textbf{Demographic:} \textit{Jewish, white, Russian }}\\
      \hline

        \small{Bengali} & \small{\textbf{Included:} \textit{dirty(2), stinky, poor(2), dangerous, smelly(2), cold, primitive, wild, tough, rich, smart, indecent, adventorous}  \textbf{Discarded:} \textit{macho, dancing, conservative, curry, rice, tiger, eastern, india, tigers(2) } \textbf{Demographic:} []}\\
       \hline

          \small{Hispanic}   & \small{\textbf{Included:} \textit{poor(3), lazy(2), dirty(2), nasty, uneducated, shady, illegal(2), aggressive, hardworking(2), beautiful, greasy, undocumented  } \textbf{Discarded:}  \textit{loud, smelly, immigrants, farmworker, spicy, spanish, gardener, season, equator  } \textbf{Demographic:} \textit{Catholic }} \\
       \hline

\hline
\hline
\multicolumn{2}{c}{\small{\textbf{Target groups from the `profession' category}}}\\
\hline
\hline

      \small{psychologist} & \small{\textbf{Included:} intelligent(3), smart(3), empathetic(3), compassionate, witty, analytical(3), educated(2), opinionated, careful, observant, curious, interesting, calm, rigid 
      \textbf{Discarded:} calming, nosy, wackjobs, helps
      \textbf{Demographic:} []} \\
       \hline
       
        \small{manager} & \small{\textbf{Included:} bossy(8), overbearing, rude, selfcentered, boss, lazy, authoritative, leader
        \textbf{Discarded: } strict(2), controlling, boring, knowledgeable, corrupt, organized(3), mean, seriousness 
        \textbf{Demographic:} white, man(2), he(2)} \\
       \hline

       \small{guitarist}  & \small{\textbf{Included:} energetic, passionate, talented(2), musical(3), young, sensitive, emotional, skilled, cool, stimulating, attitude, lazy, hippie    \textbf{Discarded:}  skinny, drunk, rock, stoned, dedication, druggies \textbf{Demographic:} male, Caucasian} \\
       \hline
       
       \small{civil servant} & \small{\textbf{Included:} lazy, tired, overworked(2), poor, bad, careless, selfish, bored, stressed, corrupt, smart, dependable  \textbf{Discarded:}  bureaucratic, kind, competent, honorable, bureaucrat(2), disrespected, democrat, state, job, volunteering
       \textbf{Demographic:} [] }\\
       \hline

       \small{butcher}  & \small{\textbf{Included:} brutal, bloody(2), violent, bloodthirsty, filthy, strong(2), sloppy, dumb, burly, macho, cuts  
       \textbf{Discarded:}  meat, fat(3), knives, foreign, uneducated(2), experienced, overweight
       \textbf{Demographic:} he(2), white, man, male} \\
       \hline
       
       \small{commander} & \small{\textbf{Included:} strong, ruthless, harsh, wise, friendly, honest, intelligent, brave, hard, trustworthy, loud(2), important, sharp, protective, strict(2), dominant    \textbf{Discarded:}  leading, experienced, leader(2), organized, bossy(2), controlling 
       \textbf{Demographic:} male(2), him}\\
       \hline
       
        \small{chess player} & \small{\textbf{Included:} smart(4), strategic(4), clever, brilliant, analytical, methodical, careful, nerdy, strategy, quiet, dorky, calm, geek  
        \textbf{Discarded:}  reserved, frail, calculated, glasses 
         \textbf{Demographic:} Asian(3), man, Russian} \\
       \hline
       
       \small{performing artist} & \small{\textbf{Included:} talented(4), charismatic, creative, skilled, egotistical(2), opinionated, attractive, expressive, outgoing, dancer 
       \textbf{Discarded:} selfish, ballerina, musical, big, emotional, ostentatious, broke  
       \textbf{Demographic:} gay} \\
       \hline
       
       \small{plumber}  & \small{\textbf{Included:} uneducated(3), lazy, illiterate, dirty(2), uncouth, fat(3), unkempt, crooked, redneck, lackadaisical, hardworking 
       \textbf{Discarded:}  greasy, expensive(2), lower, fix, installs
       \textbf{Demographic:} male, Italian}\\
       \hline
        
        \small{tailor}       & \small{\textbf{Included:} precise(2), meticulous, detailed(2), careful, professional(2), neat, skilled(2), creative, quiet(2), short 
        \textbf{Discarded:}  snarky, crafty, foreigner, james 
        \textbf{Demographic:} Italian(2), gay, Asian, man, Jew, female} \\
       \hline
       
       \small{chemist} & \small{\textbf{Included:} smart(5), intelligent(3), nerdy(3), geeky(2), studious, crafty, analytical, introverted, careful, brillant 
       \textbf{Discarded:}  shy, science, laboratory 
       \textbf{Demographic:} white, man, he}\\
       \hline
       
       \small{musician}  & \small{\textbf{Included:} creative(4), artistic(2), talented(4), professional, good, emotional, cute, intense, informal 
       \textbf{Discarded:}  emotion, poor, employed, flakey, famous, handsome, shiftless, addict, unreliable, broke 
       \textbf{Demographic:} []} \\
       \hline
       
       \small{software developer} & \small{\textbf{Included:} nerdy(7), geeky(2), smart(6), nerd, intelligent(2), genius, math, unhealthy 
       \textbf{Discarded:}  boys, rich(2), coder, glasses, skilled 
        \textbf{Demographic:} male} \\
       \hline
       
       \small{mover} & \small{\textbf{Included:} strong(18), burly, hardworking, dumb, tall 
       \textbf{Discarded:}  creepy, careless 
       \textbf{Demographic:} men(2), Mexican} \\
       \hline
       
       \small{football player}  & \small{\textbf{Included:} dumb(4), stupid(3), strong(5), tough(2), aggressive(2), fast, talented 
       \textbf{Discarded:} bully, muscular, hulking, athletic(3) injured, buff, athlete, doping 
        \textbf{Demographic:} black (2), male}\\
       \hline
       
       \small{producer} & \small{\textbf{Included:} shrewd, smart, egotistical, selfish, flashy, industrious, aloof, rich, visionary, knowledgable, businesslike, creepy, busy, critical, cheap  
       \textbf{Discarded:}  predatory, tired, leader, ideas, elitist, demanding(3), film
        \textbf{Demographic:} white, male(2)}\\
       \hline

       \small{politician} & \small{\textbf{Included:} dishonest(3), corrupt(8), deceptive, crooked, liars(3), shifty, lying(2), disingenuine, cheating, fake, slippery, argumentative 
       \textbf{Discarded:} []
       \textbf{Demographic:} female, male} \\
       \hline

       \small{prosecutor} & \small{\textbf{Included:} aggressive(5), ruthless, agressive, vindictive, harsh(2), tough, unfair, unrelenting, aggressively, sneaky, biased, contentious, persuasive, racist, corrupt, confident  
       \textbf{Discarded:}  educated, guilty(2), lawyer, justice 
       \textbf{Demographic:} male(2)}\\
       \hline

       \small{bartender} & \small{\textbf{Included:} friendly(3), sociable(2), personable, gregarious, outgoing(2), beautiful, sexy, lovely, quick, fast, stupid  
       \textbf{Discarded:}  uneducated, criminal, alcoholic(2), drunk(3), listens, drinker, provocatively, listener, listeners 
       \textbf{Demographic:} []}\\
       \hline

       \small{physicist} & \small{\textbf{Included:} smart(7), intelligent(4), nerdy(5), brilliant, eccentric, genius(2), educated, dynamic, focused   \textbf{Discarded:}  pety
       \textbf{Demographic:} man, his} \\
       \hline

       \small{assistant}  & \small{\textbf{Included:} helpful(4) efficient, diligent, organized, helper   \textbf{Discarded:}  harried, independent, meek, young, obediant, slutty, meak, unnoticed, clerical, coffee, grunt, inferior, coward 
       \textbf{Demographic:} female}\\
       \hline

      \small{historian} & \small{\textbf{Included:} bookish(4), nerdy, studious, bookworm, smart(3), academic, professorial, library   
      \textbf{Discarded:}  boring(5), dull, educated, ancient, wise, correct 
      \textbf{Demographic:} []}\\
       \hline

      \small{researcher}  & \small{\textbf{Included:} studious, nerdy(3), inquisitive, smart, analytical(2), careful, meticulous, thorough, introverted, competent, dedicated(2), awkward, shy, inquiring  \textbf{Discarded:}  boring, coldhearted, educator, overworked, intern, library, computer 
      \textbf{Demographic:} man, Asian} \\
       \hline

        \small{delivery man} & \small{\textbf{Included:} young(3), hardworking(2), poor, careless(2), stupid, teenager, strong, late(2), athletic 
        \textbf{Discarded:}  kind, fit, screwing, efficient, rushed, timely, distribute
        \textbf{Demographic:} white, black, male} \\
       \hline

       \small{entrepreneur}  & \small{\textbf{Included:} savvy(3), ambitious(2), smart, greedy(3), ruthless, young(2), creative(2), industrious, wise, rich(2), successful, inventive, talkative, sleazy, educated, extroverted    \textbf{Discarded:}  hardworker, driven, free, boss 
       \textbf{Demographic:} male}\\
       \hline

       \small{policeman} & \small{\textbf{Included:} corrupt(4), racist(2), megalomaniacal, bad, aggressive  \textbf{Discarded:}  fat, murderer, strong, safe, bully, macho, agressive, brave, hero, assaults, stern, mean, helpful
       \textbf{Demographic:} white(2)}\\
       \hline

       \small{engineer}  & \small{\textbf{Included:} smart(6), intelligent, nerdy(3), geeky, knowledgeable, egotistical, wealthy, awkward, rich, boring, intellegant, methodical 
       \textbf{Discarded:}  technical, antisocial, update 
        \textbf{Demographic:} man, Chinese, he, male} \\
       \hline

       \small{mathemati- cian} & \small{\textbf{Included:} smart(6), intelligent(3), nerdy(4), geek, analytical, analytical, nerds, good, geniuses, intelligence, meek
       \textbf{Discarded:}  logic, antisocial, introvert, numbers, algebra 
       \textbf{Demographic:} he, man(2)}\\
       \hline

      \small{nurse}  & \small{\textbf{Included:} caring(8), compassionate, hardworking, supportive, patient, kind, dedicated 
      \textbf{Discarded:}  overworked, profession, nice, busy, tired, hot, underqualified
      \textbf{Demographic:} woman(2), her, female(2), she, male}\\
       \hline

       \small{prisoner}  & \small{\textbf{Included:} violent(5), dangerous(2), brutal, cruel, evil(2), criminal, bad, dishonest, untrustworthy, hopeless, thug, lazy, perpetrator 
       \textbf{Discarded:} smelly, guilty(2), mean(2)  
        \textbf{Demographic:} black(6)}\\
       \hline

           \caption{Target groups and associated stereotype words in StereoSet. Words which occur more than once for a given group have their frequency indicated in parentheses. Words that are included in the analysis are ranked by closeness to the cluster mean; thus the first words in the list are most representative of the stereotype for that group. }
    \label{tab:stereoset-data}
\end{longtable}
\clearpage
\twocolumn


\end{document}